\documentclass[11pt,superscriptaddress,aps,prd,preprint,showpacs]{revtex4}

\usepackage[dvips]{graphicx}
\usepackage{amsfonts}
\usepackage{slashed}
\usepackage[utf8x]{inputenc}
\usepackage{amsmath}
\usepackage{hyperref}
\usepackage{cleveref}
\usepackage{xcolor}

\begin{document}

\title{On perturbative aspects of a nonminimal Lorentz-violating QED with CPT-odd dimension-6 terms}

\author{T. Mariz}
\affiliation{Instituto de F\'\i sica, Universidade Federal de Alagoas,\\ 57072-900, Macei\'o, Alagoas, Brazil}
\email{tmariz,max.melo@fis.ufal.br}

\author{M. Melo}
\affiliation{Instituto de F\'\i sica, Universidade Federal de Alagoas,\\ 57072-900, Macei\'o, Alagoas, Brazil}
\email{tmariz, max.melo@fis.ufal.br}

\author{J. R. Nascimento}
\affiliation{Departamento de F\'\i sica, Universidade Federal da Para\'\i ba\\ Caixa Postal 5008, 58051-970, Jo\~ao Pessoa, Para\'\i ba, Brazil}
\email{jroberto, petrov@fisica.ufpb.br}

\author{A. Yu. Petrov}
\affiliation{Departamento de F\'\i sica, Universidade Federal da Para\'\i ba\\ Caixa Postal 5008, 58051-970, Jo\~ao Pessoa, Para\'\i ba, Brazil}
\email{jroberto, petrov@fisica.ufpb.br}

\begin{abstract}
In this paper, we investigate the first-order one-loop quantum corrections generated by all possible nonminimal dimension-6 additive CPT-odd Lorentz-violating (LV) terms to analyze the radiative induction of the Carroll–Field–Jackiw (CFJ)-like terms. We demonstrate that for dimension-6 operators, the resulting CFJ term can diverge for some choices of LV operators; however, in all these cases, this divergence is eliminated for certain relations between LV parameters.
\end{abstract}


\maketitle

\section{Introducion}

In recent decades, the study of effective field theories has gained prominence in the search for possible signals of a more fundamental underlying physics. In this context, Kosteleck\'y and collaborators \cite{Kostelecky:2003fs,Colladay:1998fq} developed the Standard-Model Extension (SME), which serves as the paradigmatic theoretical framework to describe such violations. The SME preserves the essential features of the conventional Standard Model, such as field content and gauge symmetries, while incorporating small corrective terms that can be interpreted as remnants of physics originating at the Planck energy scale. The terms introduced to represent LV effects typically contain constant background tensors, referred to as LV coefficients, contracted with local tensor operators.

Within the framework of the SME, quantum electrodynamics (QED) can be generalized by including terms that violate Lorentz symmetry in the fermionic sector. In addition to minimal extensions, which involve operators with mass dimension $d \leq 4$  (for a detailed discussion of the minimal LV extension of QED, see \cite{Kostelecky:2001jc}, one can also consider nonminimal terms involving higher-dimensional operators, which, with dimensions up to 6 in the fermion-gauge coupling sector and the pure fermionic sector and up to 8 in the pure gauge sector, have been listed in \cite{Kostelecky:2018yfa}, see also \cite{Ding:2016lwt} (a general discussion of higher-dimensional LV operators in spinor QED context is presented in  \cite{Kostelecky:2009zp,Kostelecky:2013rta}). It should be noted that studies of higher-dimensional operators, either in Lorentz-invariant or LV cases, present an important line in investigations of extensions of the Standard Model; see, e.g., \cite{Grzadkowski:2010es}. Moreover, terms with dimensions up to eight are considered in this context \cite{Li:2020gnx}. Further, various impacts of such operators, including perturbative ones which will be the main ones in this paper, have been studied -- among important results, one can emphasize explicit obtaining of one-loop corrections generated by dimension-5 operators, both CPT-odd \cite{dim51} and CPT-even \cite{dim52} ones, phenomenological estimations for the values of nonminimal LV parameters \cite{Araujo:2015zsa}, study of ambiguities of quantum corrections generated by such operators \cite{Fargnoli:2025buj}, of modification of dispersion relations \cite{Reyes:2014wna}, and of impacts of such operators in the Cherenkov radiation context \cite{Petrov:2025wey}. It is worth mentioning that one of the first known examples of perturbative generating the CPT-even term has been performed with use of the nonminimal coupling \cite{aether}. Therefore, the study of quantum impacts of LV dimension-6 operators, which, up to now, have been studied only within the tree-level context see \cite{Casana:2018rhg,Araujo:2019txk,Araujo:2019sdg}, seems to be rather natural.

Following the formulation presented in \cite{Kostelecky:2018yfa}, this work focuses on the one-loop quantum corrections induced by all CPT-odd QED operators of dimension-6, to investigate the radiative generation of the CFJ term, introduced originally in \cite{CFJ}, as well as its higher-derivative analogues. Earlier, various issues related to the emergence of this term as a quantum correction were studied; see e.g. \cite{Colladay:1998fq,Jackiw:1999qq,Andrianov:2001zj,Chung:1998jv,Chung:1999pt,Mariz:2005jh} and many other papers.

The structure of the paper looks as follows. In Section 2, we write down our CPT-odd LV extension of QED. In Section 3, we calculate the leading-order one-loop corrections via the derivative expansion method. In Section 4, we apply an alternative method of calculation based on the Feynman parametrization. Our results are summarized in Section 5.

\section{The QED model with dimension $d=6$ operators}

To begin our study, let us write down the Lagrangian for extended QED, which includes possible dimension-6 LV operators, while restricting ourselves exclusively to those that violate CPT symmetry, i.e., CPT-odd terms. All these terms are listed in \cite{Kostelecky:2018yfa}. To this end, we consider the Lagrangian, which looks like:
\begin{eqnarray}\label{lagrangiana1}
    {\cal L}_{\psi} &=& \bar\psi(i\slashed{D}-m)\psi + \frac{1}{2} e^{(6)\alpha\beta\gamma}\bar\psi i D_{(\alpha}iD_{\beta}iD_{\gamma)}\psi + h.c +\frac{1}{2}i f^{(6)\alpha\beta\gamma}\bar\psi \gamma_{5} i D_{(\alpha}iD_{\beta}iD_{\gamma)}\psi +h.c.   \nonumber\\
    &+& \frac{1}{4}g^{(6)\mu\nu\alpha\beta\gamma}\bar\psi\sigma_{\mu\nu} i D_{(\alpha}iD_{\beta}iD_{\gamma)}\psi + h.c. +\frac{1}{4}e_{F}^{(6)\alpha\beta\gamma}\bar\psi F_{\beta\gamma}iD_{\alpha}\psi+h.c.\nonumber\\
    &+& \frac{1}{4}if_{F}^{(6)\alpha\beta\gamma}\bar\psi \gamma_{5}F_{\beta\gamma}iD_{\alpha}\psi+h.c. 
    +  \frac{1}{8}g_{F}^{(6)\mu\nu\alpha\beta\gamma}\bar\psi \sigma_{\mu\nu}F_{\beta\gamma}iD_{\alpha}\psi + h.c\nonumber\\
    &-&\frac{1}{2}m_{DF}^{(6)\alpha\beta\gamma}\bar\psi(D_{\alpha}F_{\beta\gamma})\psi - \frac{1}{2}im_{5DF}^{(6)\alpha\beta\gamma}\bar\psi\gamma_{5}(D_{\alpha}F_{\beta\gamma})\psi - \frac{1}{4}H_{DF}^{\mu\nu\alpha\beta\gamma}\bar\psi\sigma_{\mu\nu}(D_{\alpha}F_{\beta\gamma})\psi. 
\end{eqnarray}
Here, the parameters $e^{(6)\alpha\beta\gamma}$, $f^{(6)\alpha\beta\gamma}$, $g^{(6)\mu\nu\alpha\beta\gamma}$, etc., are constants. As usual, the gauge covariant derivative is defined as $D_{\mu} = \partial_{\mu} + ieA_{\mu}$, so that it yields
 \begin{eqnarray}
    i D_{(\alpha} i D_{\beta} i D_{\gamma)} \psi &=& \frac{1}{6} (i D_{\alpha} i D_{\beta} i D_{\gamma} + i D_{\alpha} i D_{\gamma} i D_{\beta} + i D_{\beta} i D_{\alpha} i D_{\gamma} \nonumber \\
    && + i D_{\beta} i D_{\gamma} i D_{\alpha} + i D_{\gamma} i D_{\alpha} i D_{\beta} + i D_{\gamma} i D_{\beta} i D_{\alpha})\psi
\end{eqnarray}
or, as is the same,
\begin{eqnarray}
    i D_{(\alpha}i D_{\beta}i D_{\gamma)}\psi &=& i\partial_{\alpha}\partial_{\beta}\partial_{\gamma}\psi + e\left[ (\partial_{\alpha}\partial_{\beta}A_{\gamma}) + 3(\partial_{\alpha} A_{\beta})\partial_{\gamma} + 3 A_{\alpha}\partial_{\beta}\partial_{\gamma}\right]\psi\nonumber\\
    &+&ie^2 \left[ (\partial_{\alpha}A_{\beta})A_{\gamma} + 2 A_{\alpha}(\partial_{\beta}A_{\gamma}) + 2 A_{\alpha}A_{\beta}\partial_{\gamma}\right]\psi - e^3 A_{\alpha}A_{\beta}A_{\gamma}\psi.
\end{eqnarray}

In this way, writing the covariant derivative explicitly, one arrives at the following form of the Lagrangian \eqref{lagrangiana1}:
\begin{eqnarray}\label{lagrangiana2}
    {\cal L}_{\psi} &=&\bar\psi [ i \slashed{\partial} - e\slashed{A} - m - i(e^{(6)\alpha\beta\gamma}+i f^{(6)\alpha\beta\gamma}\gamma_{5}+\frac{1}{2}g^{(6)\mu\nu\alpha\beta\gamma}\sigma_{\mu\nu})\partial_{\alpha}\partial_{\beta}\partial_{\gamma} \nonumber\\ 
    &&+e(e^{(6)\alpha\beta\gamma}+i f^{(6)\alpha\beta\gamma}\gamma_{5}+\frac{1}{2}g^{(6)\mu\nu\alpha\beta\gamma}\sigma_{\mu\nu})\nabla_{\alpha\beta}A_{\gamma}\nonumber\\
    &&+ ie^2(e^{(6)\alpha\beta\gamma}+i f^{(6)\alpha\beta\gamma}\gamma_{5}+\frac{1}{2}g^{(6)\mu\nu\alpha\beta\gamma}\sigma_{\mu\nu})\nabla_{\alpha}A_{\beta}A_{\gamma}\nonumber\\
    && - e^3(e^{(6)\alpha\beta\gamma}+i f^{(6)\alpha\beta\gamma}\gamma_{5}+\frac{1}{2}g^{(6)\mu\nu\alpha\beta\gamma}\sigma_{\mu\nu})A_{\alpha}A_{\beta}A_{\gamma}\nonumber\\
    &&+ ie ( e_{F}^{(6)\alpha\beta\gamma} + if_{F}^{(6)\alpha\beta\gamma}\gamma_{5} + \frac{1}{2}g_{F}^{(6)\mu\nu\alpha\beta\gamma}\sigma_{\mu\nu})(\partial_{\beta}A_{\gamma})\partial_{\alpha}\nonumber\\
    && - e^2( e_{F}^{(6)\alpha\beta\gamma} + if_{F}^{(6)\alpha\beta\gamma}\gamma_{5} + \frac{1}{2}g_{F}^{(6)\mu\nu\alpha\beta\gamma}\sigma_{\mu\nu})(\partial_{\beta}A_{\gamma})A_{\alpha}\nonumber \\
    &&- e(m_{DF}^{(6)\alpha\beta\gamma} +im_{5DF}^{(6)\alpha\beta\gamma}\gamma_{5}+\frac{1}{2}H_{DF}^{(6)\mu\nu\alpha\beta\gamma}\sigma_{\mu\nu})\partial_{\alpha}(\partial_{\beta}A_{\gamma})\nonumber\\
    &&+ ie^2(m_{DF}^{(6)\alpha\beta\gamma} + im_{5DF}^{(6)\alpha\beta\gamma}\gamma_{5}+\frac{1}{2}H_{DF}^{(6)\mu\nu\alpha\beta\gamma}\sigma_{\mu\nu})A_{\alpha}\partial_{\beta}A_{\gamma}]\psi, 
\end{eqnarray}
where, in Eq.~\eqref{lagrangiana2}, we have adopted the convention
$F_{\mu\nu} \to e F_{\mu\nu}$ and employed the notation
\begin{eqnarray}
    &&\nabla_{\alpha\beta}A_{\gamma}\psi = \left( \partial_{\alpha}\partial_{\beta}A_{\gamma}\right)\psi +  3(\partial_{\alpha}A_{\beta})\partial_{\gamma}\psi + 3A_{\alpha}\partial_{\beta}\partial_{\gamma}\psi,  \nonumber\\
    && \nabla_{\alpha}A_{\beta}A_{\gamma}\psi=(\partial_{\alpha}A_{\beta})A_{\gamma}\psi + 2A_{\alpha}(\partial_{\beta}A_{\gamma})\psi + 3A_{\alpha}A_{\beta}\partial_{\gamma}\psi.
\end{eqnarray}

We emphasize that the coefficients $e^{(6)\alpha\beta\gamma}$, $f^{(6)\alpha\beta\gamma}$, and $g^{(6)\mu\nu\alpha\beta\gamma}$ are symmetric in the indices $\alpha,\beta,\gamma$, whereas the coefficients $e_{F}^{(6)\alpha\beta\gamma}$, $f_{F}^{(6)\alpha\beta\gamma}$, $g_{F}^{(6)\mu\nu\alpha\beta\gamma}$, $m_{DF}^{(6)\alpha\beta\gamma}$, $m_{5DF}^{(6)\alpha\beta\gamma}$, and $H_{DF}^{(6)\mu\nu\alpha\beta\gamma}$ are antisymmetric in the indices $\beta, \gamma$. The indices $\mu,\nu$ of all coefficients are always antisymmetric because they are contracted with the sigma matrix $\sigma_{\mu\nu}$.

To compute the one-loop effective action $S_{eff}$ (the fermionic determinant), we will need to use the fermionic generating functional
\begin{eqnarray}\label{funcionalgerador}
    Z = \int D\bar\psi D\psi e^{i\int d^4x{\cal L}_{\psi}} = e^{iS_{eff}}. 
\end{eqnarray}
Thus, by considering the Lagrangian~\eqref{lagrangiana2} and integrating out the fermions, we obtain the effective action
\begin{eqnarray}\label{Seff}
    S_{eff} &=& -i\mathrm{Tr} \ln [ \slashed{p} - e\slashed{A} - m + (e^{(6)\alpha\beta\gamma}+i f^{(6)\alpha\beta\gamma}\gamma_{5}+\frac{1}{2}g^{(6)\mu\nu\alpha\beta\gamma}\sigma_{\mu\nu})p_{\alpha}p_{\beta}p_{\gamma} \nonumber\\ 
    &&-e(e^{(6)\alpha\beta\gamma}+i f^{(6)\alpha\beta\gamma}\gamma_{5}+\frac{1}{2}g^{(6)\mu\nu\alpha\beta\gamma}\sigma_{\mu\nu})\nabla_{\alpha\beta}A_{\gamma}\nonumber\\
    &&+e^2(e^{(6)\alpha\beta\gamma}+i f^{(6)\alpha\beta\gamma}\gamma_{5}+\frac{1}{2}g^{(6)\mu\nu\alpha\beta\gamma}\sigma_{\mu\nu})\nabla_{\alpha}A_{\beta}A_{\gamma}\nonumber\\
    && - e^3(e^{(6)\alpha\beta\gamma}+i f^{(6)\alpha\beta\gamma}\gamma_{5}+\frac{1}{2}g^{(6)\mu\nu\alpha\beta\gamma}\sigma_{\mu\nu})A_{\alpha}A_{\beta}A_{\gamma}\nonumber\\
    &&-ie ( e_{F}^{(6)\alpha\beta\gamma} + if_{F}^{(6)\alpha\beta\gamma}\gamma_{5} + \frac{1}{2}g_{F}^{(6)\mu\nu\alpha\beta\gamma}\sigma_{\mu\nu})(k_{\beta}A_{\gamma})p_{\alpha}\nonumber\\
    && +ie^2( e_{F}^{(6)\alpha\beta\gamma} + if_{F}^{(6)\alpha\beta\gamma}\gamma_{5} + \frac{1}{2}g_{F}^{(6)\mu\nu\alpha\beta\gamma}\sigma_{\mu\nu})(k_{\beta}A_{\gamma})A_{\alpha}\nonumber \\
    &&+e (m_{DF}^{(6)\alpha\beta\gamma} + im_{5DF}^{(6)\alpha\beta\gamma}\gamma_{5}+\frac{1}{2}H_{DF}^{(6)\mu\nu\alpha\beta\gamma}\sigma_{\mu\nu})k_{\alpha}k_{\beta}A_{\gamma}\nonumber\\
    &&-e^2(m_{DF}^{(6)\alpha\beta\gamma} + im_{5DF}^{(6)\alpha\beta\gamma}\gamma_{5}+\frac{1}{2}H_{DF}^{(6)\mu\nu\alpha\beta\gamma}\sigma_{\mu\nu})A_{\alpha}k_{\beta}A_{\gamma}].
\end{eqnarray}
We now aim to expand Eq.~\eqref{Seff} in a power series in the external fields, so that we can write $S_{\text{eff}}$ as
\begin{eqnarray}
    S_{eff} &=& S_{eff}^{(0)} + \sum_{n=1}^{\infty} S_{eff}^{(n)},
\end{eqnarray}
where $S_{eff}^{(0)} = -i\mathrm{Tr}\ln G^{-1}(p)$ and
\begin{eqnarray}
\label{seffn}
S_{eff}^{(n)} &=& \frac{i}{n} \, \mathrm{Tr} \{G(p)[e\slashed{A} 
+ e ( e^{(6)\alpha\beta\gamma} + i f^{(6)\alpha\beta\gamma} \gamma_5 
+ \textstyle{\frac{1}{2}} g^{(6)\mu\nu\alpha\beta\gamma} \sigma_{\mu\nu} ) 
\nabla_{\alpha\beta}(p,k) A_\gamma \nonumber \\
&& - e^2 ( e^{(6)\alpha\beta\gamma} + i f^{(6)\alpha\beta\gamma} \gamma_5 
+ \textstyle{\frac{1}{2}} g^{(6)\mu\nu\alpha\beta\gamma} \sigma_{\mu\nu} ) 
\nabla_{\alpha}(p,k) A_\beta A_\gamma \nonumber \\
&& + e^3 ( e^{(6)\alpha\beta\gamma} + i f^{(6)\alpha\beta\gamma} \gamma_5 
+ \textstyle{\frac{1}{2}} g^{(6)\mu\nu\alpha\beta\gamma} \sigma_{\mu\nu} ) 
A_\alpha A_\beta A_\gamma \nonumber \\
&& + ie ( e_F^{(6)\alpha\beta\gamma} + i f_F^{(6)\alpha\beta\gamma} \gamma_5 
+ \textstyle{\frac{1}{2}} g_F^{(6)\mu\nu\alpha\beta\gamma} \sigma_{\mu\nu} ) 
(k_\beta A_\gamma) p_\alpha \nonumber \\
&& - i e^2 ( e_F^{(6)\alpha\beta\gamma} + i f_F^{(6)\alpha\beta\gamma} \gamma_5 
+ \textstyle{\frac{1}{2}} g_F^{(6)\mu\nu\alpha\beta\gamma} \sigma_{\mu\nu} ) 
(k_\beta A_\gamma) A_\alpha \nonumber \\
&& -e ( m_{DF}^{(6)\alpha\beta\gamma} + i m_{5DF}^{(6)\alpha\beta\gamma} \gamma_5 
+ \textstyle{\frac{1}{2}} H_{DF}^{(6)\mu\nu\alpha\beta\gamma} \sigma_{\mu\nu} ) 
k_\alpha k_\beta A_\gamma \nonumber \\
&& + e^2 ( m_{DF}^{(6)\alpha\beta\gamma} + i m_{5DF}^{(6)\alpha\beta\gamma} \gamma_5 
+ \textstyle{\frac{1}{2}} H_{DF}^{(6)\mu\nu\alpha\beta\gamma} \sigma_{\mu\nu} ) 
A_\alpha k_\beta A_\gamma]\}^n,
\label{Seff^n}
\end{eqnarray}
with
\begin{eqnarray}
    G(p) &=& \frac{1}{\slashed{p} - m  + \left(e^{(6)\alpha\beta\gamma}+i f^{(6)\alpha\beta\gamma}\gamma_{5}+\textstyle{\frac{1}{2}}g^{(6)\mu\nu\alpha\beta\gamma}\sigma_{\mu\nu}\right)p_{\alpha}p_{\beta}p_{\gamma} },
\end{eqnarray}
$\nabla_{\alpha\beta}(p,k) =k_{\alpha}k_{\beta} + 3k_{\alpha}p_{\beta} + 3p_{\alpha}p_{\beta}$, and $\nabla_{\alpha}(p,k)=3k_{\alpha} + 3p_{\alpha}$.

At this point, we begin to calculate the contributions to (\ref{seffn}) explicitly. To do this, we perform the trace over the coordinate space and make use of the commutation relation $A_{\mu}(x)G(p)=G(p-k)A_{\mu}(x)$, which holds for the operator $G(p)$. From a formal point of view, this relation establishes the basis for the derivative expansion framework described in detail in \cite{Aitchison:1984ys,Aitchison:1985pp}. Relevant contributions are those that are quadratic in vector fields and linear in coefficients associated with operators. For $n=1$, these contributions are
\begin{eqnarray}
    S_{eff}^{(1)} = i \int d^4x ( \Pi_{1a}^{\gamma\delta} +\Pi_{1b}^{\gamma\delta} + \Pi_{1c}^{\gamma\delta} )A_{\gamma}A_{\delta},
\end{eqnarray}
where
\begin{subequations}
\begin{eqnarray}
    \Pi_{1a}^{\gamma\delta} &=& -e^2\int \frac{d^4p}{(2\pi)^4}\mathrm{tr}\ G(p)(e^{(6)\alpha\gamma\delta}+i f^{(6)\alpha\gamma\delta}\gamma_{5}+\textstyle{\frac{1}{2}}g^{(6)\mu\nu\alpha\gamma\delta}\sigma_{\mu\nu})\nabla_{\alpha}(p,k),\\
    \Pi_{1b}^{\gamma\delta} &=& -ie^2\int \frac{d^4p}{(2\pi)^4}\mathrm{tr}\ G(p)(e_{F}^{(6)\gamma\beta\delta}+i f_{F}^{(6)\gamma\beta\delta}\gamma_{5}+\textstyle{\frac{1}{2}}g_{F}^{(6)\mu\nu\gamma\beta\delta}\sigma_{\mu\nu})k_{\beta},\\
    \Pi_{1c}^{\gamma\delta} &=& e^2\int \frac{d^4p}{(2\pi)^4}\mathrm{tr}\ G(p)(m_{DF}^{(6)\gamma\beta\delta} + im_{5DF}^{(6)\gamma\beta\delta}\gamma_{5}+\textstyle{\frac{1}{2}}H_{DF}^{(6)\mu\nu\gamma\beta\delta}\sigma_{\mu\nu})k_{\beta}.
\end{eqnarray}
\end{subequations}
Now, using the expansion of the propagator $G(p)$ in series in LV parameters, given by
\begin{eqnarray}\label{Gexp}
    G(p) = S(p) - S(p) (e^{(6)\alpha\beta\gamma}+i f^{(6)\alpha\beta\gamma}\gamma_{5}+\textstyle{\frac{1}{2}}g^{(6)\mu\nu\alpha\beta\gamma}\sigma_{\mu\nu})p_{\alpha}p_{\beta}p_{\gamma}S(p) + \cdots,
\end{eqnarray}
with $S(p) = (\slashed{p} - m)^{-1}$, the first-order contributions in the coefficients take the form:
\begin{subequations}\label{pi1}
\begin{align}
\Pi_{1a}^{\gamma\delta} &=
- e^2 \int \frac{d^4p}{(2\pi)^4}
\mathrm{tr}\ S(p)
\left(
e^{(6)\alpha\gamma\delta}
+ i f^{(6)\alpha\gamma\delta}\gamma_{5}
+ \frac{1}{2}g^{(6)\mu\nu\alpha\gamma\delta}\sigma_{\mu\nu}
\right)
\nabla_{\alpha}(p,k),
\label{pi1a}
\\
\Pi_{1b}^{\gamma\delta} &=
- i e^2 \int \frac{d^4p}{(2\pi)^4}
\mathrm{tr}\ S(p)
\left(
e_{F}^{(6)\gamma\beta\delta}
+ i f_{F}^{(6)\gamma\beta\delta}\gamma_{5}
+ \frac{1}{2}g_{F}^{(6)\mu\nu\gamma\beta\delta}\sigma_{\mu\nu}
\right)
k_{\beta},
\label{pi1b}
\\
\Pi_{1c}^{\gamma\delta} &=
e^2 \int \frac{d^4p}{(2\pi)^4}
\mathrm{tr}\ S(p)
\left(
m_{DF}^{(6)\gamma\beta\delta}
+ i m_{5DF}^{(6)\gamma\beta\delta}\gamma_{5}
+ \frac{1}{2}H_{DF}^{(6)\mu\nu\gamma\beta\delta}\sigma_{\mu\nu}
\right)
k_{\beta}.
\label{pi1c}
\end{align}
\end{subequations}
The contributions for $n = 2$ in Eq.~\eqref{Seff^n} are as follows:
\begin{eqnarray}
        S_{eff}^{(2)} = \frac{i}{2}\int d^4x (\Pi_{2a}^{\gamma\delta} + \Pi_{2b}^{\gamma\delta}+\Pi_{2c}^{\gamma\delta}+\Pi_{2d}^{\gamma\delta}+\Pi_{2e}^{\gamma\delta}+\Pi_{2f}^{\gamma\delta}+\Pi_{2g}^{\gamma\delta}+\Pi_{2h}^{\gamma\delta})A_{\gamma}A_{\delta},
\end{eqnarray}
where
\begin{subequations}\label{Pi2}
\begin{eqnarray}
    \label{pi2a}\Pi_{2a}^{\gamma\delta} &=& -e^2\int \frac{d^4p}{(2\pi)^4} \mathrm{tr}\ S(p)\gamma^{\gamma}S(p-k)(e^{(6)\alpha\beta\kappa}+i f^{(6)\alpha\beta\kappa}\gamma_{5}+\frac{1}{2}g^{(6)\mu\nu\alpha\beta\kappa}\sigma_{\mu\nu})\\
    &&\times(p-k)_{\alpha}(p-k)_{\beta}(p-k)_{\kappa}S(p-k)\gamma^{\delta},\nonumber\\
    \label{pi2b}\Pi_{2b}^{\gamma\delta} &=& -e^2\int \frac{d^4p}{(2\pi)^4}\mathrm{tr}\ S(p) (e^{(6)\alpha\beta\kappa}+i f^{(6)\alpha\beta\kappa}\gamma_{5}+\frac{1}{2}g^{(6)\mu\nu\alpha\beta\kappa}\sigma_{\mu\nu})p_{\alpha}p_{\beta}p_{\kappa}S(p)\gamma^{\gamma}S(p-k)\gamma^{\delta},\hspace{0.7cm} \\
    \label{pi2c}\Pi_{2c}^{\gamma\delta} &=& e^2\int \frac{d^4p}{(2\pi)^4}\mathrm{tr}\ S(p) (e^{(6)\alpha\beta\gamma}+i f^{(6)\alpha\beta\gamma}\gamma_{5}+\frac{1}{2}g^{(6)\mu\nu\alpha\beta\gamma}\sigma_{\mu\nu})\nabla_{\alpha\beta}(p,k)S(p-k)\gamma^{\delta},\\
    \label{pi2d}\Pi_{2d}^{\gamma\delta} &=& e^2\int \frac{d^4p}{(2\pi)^4} \mathrm{tr}\ S(p)\gamma^{\gamma}S(p-k) (e^{(6)\alpha\beta\delta}+i f^{(6)\alpha\beta\delta}\gamma_{5}+\frac{1}{2}g^{(6)\mu\nu\alpha\beta\delta}\sigma_{\mu\nu})\nabla_{\alpha\beta}(p-k, -k),\\
    \label{pi2e}\Pi_{2e}^{\gamma\delta} &=& ie^2\int \frac{d^4p}{(2\pi)^4} \mathrm{tr}\ S(p)\gamma^{\gamma}S(p-k) (e_{F}^{(6)\alpha\beta\delta}+i f_{F}^{(6)\alpha\beta\delta}\gamma_{5}+\frac{1}{2}g_{F}^{(6)\mu\nu\alpha\beta\delta}\sigma_{\mu\nu})(p-k)_{\alpha}(-k_{\beta}),\\
    \label{pi2f}\Pi_{2f}^{\gamma\delta} &=& ie^2\int \frac{d^4p}{(2\pi)^4}\mathrm{tr}\ S(p) (e_{F}^{(6)\alpha\beta\gamma}+i f_{F}^{(6)\alpha\beta\gamma}\gamma_{5}+\frac{1}{2}g_{F}^{(6)\mu\nu\alpha\beta\gamma}\sigma_{\mu\nu})p_{\alpha}k_{\beta}S(p-k)\gamma^{\delta} ,\\
    \label{pi2g}\Pi_{2g}^{\gamma\delta} &=& -e^2\int \frac{d^4p}{(2\pi)^4} \mathrm{tr}\ S(p)\gamma^{\gamma}S(p-k)(m_{DF}^{(6)\alpha\beta\delta} + im_{5DF}^{(6)\alpha\beta\delta}\gamma_{5}+\frac{1}{2}H_{DF}^{(6)\mu\nu\alpha\beta\delta}\sigma_{\mu\nu})k_{\alpha}k_{\beta},\\
    \label{pi2h}\Pi_{2h}^{\gamma\delta} &=& -e^2\int \frac{d^4p}{(2\pi)^4} \mathrm{tr}\ S(p)(m_{DF}^{(6)\alpha\beta\gamma} + im_{5DF}^{(6)\alpha\beta\gamma}\gamma_{5}+\frac{1}{2}H_{DF}^{(6)\mu\nu\alpha\beta\gamma}\sigma_{\mu\nu})k_{\alpha}k_{\beta}S(p-k)\gamma^{\delta}.
\end{eqnarray}
\end{subequations}
Note that, once more, we have employed the propagator expansion (\ref{Gexp}) to keep only the first-order terms in the coefficients.

Due to the symmetry of the coefficient $e^{(6)\alpha\beta\gamma}$ in its indices, it is not possible to induce the CFJ term with its use. The generation of such a term requires a coefficient structure involving a contraction with the Levi-Civita tensor $\epsilon^{\mu\nu\sigma\kappa}$ and a constant axial vector, which is incompatible with the symmetry of $e^{(6)\alpha\beta\gamma}$. 

In the next section, we will investigate the induction of the CFJ term through the derivative expansion method.

\section{Derivative Expansion}

To generate the CFJ term using the derivative expansion method, it is sufficient to restrict ourselves to terms containing only one derivative.
We start by analyzing the contributions corresponding to $n=1$, namely the amplitudes associated with the equations \eqref{pi1a}, \eqref{pi1b} and \eqref{pi1c}.

Thus, only the coefficients $e_{F}^{(6)\alpha\beta\gamma}$ and $m_{DF}^{(6)\alpha\beta\gamma}$ yield non-vanishing contributions, while the remaining ones vanish upon evaluation of the traces, so that $\Pi^{\gamma\delta}_{1a}\rightarrow\Pi^{\gamma\delta}_{CFJ,1a}=0$. We therefore retain only the contributions $\Pi^{\gamma\delta}_{1b}\rightarrow\Pi^{\gamma\delta}_{CFJ,1b}$ and $\Pi^{\gamma\delta}_{1c}\rightarrow\Pi^{\gamma\delta}_{CFJ,1c}$, which take the form
\begin{subequations}
\begin{align}
\Pi^{\gamma\delta}_{CFJ,1b} &=
- i e^2 \int \frac{d^4p}{(2\pi)^4}
\mathrm{tr}\ S(p)
e_{F}^{(6)\gamma\beta\delta}
k_{\beta},
\label{pi1b,ef}
\\
\Pi^{\gamma\delta}_{CFJ,1c} &=
e^2 \int \frac{d^4p}{(2\pi)^4}
\mathrm{tr}\ S(p)
m_{DF}^{(6)\gamma\beta\delta}
k_{\beta}.
\label{pi1c,ef}
\end{align}
\end{subequations}

Power counting of the above integrals reveals that they are superficially divergent (nevertheless, we note that the same situation occurs within various schemes of calculating the CFJ term, including minimal ones, and this fact explains its ambiguity; see e.g. \cite{Colladay:1998fq,dim51,Jackiw:1999qq,Mariz:2005jh} and many other papers). To address this issue, we use the method of dimensional regularization, for which we need to make the replacement
\begin{eqnarray}
    \int \frac{d^4p}{(2\pi)^4} \rightarrow \mu^{4-D}\int \frac{d^Dp}{(2\pi)^D},
\end{eqnarray}
which yields the following expressions:
\begin{subequations}
    \begin{eqnarray}
    \Pi^{\gamma\delta}_{CFJ,1b} &=& - 4ie^2m \mu^{4-D} \int \frac{d^Dp}{(2\pi)^D}\frac{1}{(p^2-m^2)}k_{\beta}\delta^{\gamma}_{\ \kappa}\delta^{\delta}_{\ \lambda}e_{F}^{(6)\kappa\beta\lambda},\\
    \Pi^{\gamma\delta}_{CFJ,1c} &=& 4e^2m\mu^{4-D} \int \frac{d^Dp}{(2\pi)^D}\frac{1}{(p^2-m^2)}k_{\beta}\delta^{\gamma}_{\ \kappa}\delta^{\delta}_{\ \lambda}m_{DF}^{(6)\kappa\beta\lambda}.
\end{eqnarray}
\end{subequations}
With this, after performing the integrations, the results can be written as
\begin{subequations}
    \begin{eqnarray}
    \Pi^{\gamma\delta}_{CFJ,1b} &=& - \frac{2^{2-D} \pi ^{1-\frac{D}{2}} e^2 \mu ^{4-D} m^{D-1}  \csc (\pi  D)   }{ \Gamma \left(\frac{D}{2}\right)}k_{\beta}\delta^{\gamma}_{\ \kappa}\delta^{\delta}_{\ \lambda}e_{F}^{(6)\kappa\beta\lambda},\\
     \Pi^{\gamma\delta}_{CFJ,1c} &=& - \frac{i 2^{2-D} \pi ^{1-\frac{D}{2}} e^2 \mu ^{4-D} m^{D-1}  \csc (\pi  D)  }{ \Gamma \left(\frac{D}{2}\right)}m_{DF}^{(6)\kappa\beta\delta}k_{\beta } \delta^{\gamma}_{\ \kappa }.
\end{eqnarray}
\end{subequations}

We now perform a series expansion of $D\rightarrow4$ and make use of the fact that the LV coefficients can be expressed as 
\begin{eqnarray}\label{coeficiente_eFm_DF}
\label{eF}e_{F}^{(6)\alpha\beta\gamma} &=& \epsilon^{\alpha\beta\gamma\lambda}\,(e_{F})_{\lambda},\\
\label{mDF}m_{DF}^{(6)\alpha\beta\gamma} &=& \epsilon^{\alpha\beta\gamma\lambda}\,(m_{DF})_{\lambda},
\end{eqnarray}
where we restrict ourselves to the case where $e_{F}^{(6)\alpha\beta\gamma}$ and $m_{DF}^{(6)\alpha\beta\gamma}$ are totally antisymmetric with respect to their indices. Using these expressions, we arrive at the results
\begin{subequations}
\begin{eqnarray}
    \Pi^{\gamma\delta}_{CFJ,1b} &=& -\frac{e^2m^3}{2\pi^2\epsilon^{\prime}}\epsilon^{\gamma\delta\alpha\beta}k_{\alpha}(e_{F})_{\beta} -\frac{e^2m^3}{4\pi^2}\epsilon^{\gamma\delta\alpha\beta}k_{\alpha}(e_{F})_{\beta},\\
    \label{CFJ1c}\Pi^{\gamma\delta}_{CFJ,1c}
    &=& -\frac{ie^2m^3}{2\pi^2\epsilon^{\prime}}\epsilon^{\gamma\delta\alpha\beta}k_{\alpha}(m_{DF})_{\beta} - \frac{ie^2m^3}{4\pi^2}\epsilon^{\gamma\delta\alpha\beta}k_{\alpha}(m_{DF})_{\beta},
\end{eqnarray}
\end{subequations}
where $\frac{1}{\epsilon^{\prime}} = \frac{1}{\epsilon} - \ln \frac{m}{\mu^{\prime}}$, with $\epsilon = 4-D$ and $\mu^{\prime} = 4\pi\mu^{2}\epsilon^{-\gamma}$.

We next evaluate the contributions corresponding to $n = 2$ in Eq.~(\ref{Pi2}). To isolate the single-derivative contributions, we adopt the expansion
\begin{eqnarray}
    S(p-k) = S(p) + S(p)\slashed{k}S(p) + \cdots.
\end{eqnarray}
Thus, for the amplitudes~(\ref{pi2a}), (\ref{pi2b}), (\ref{pi2c}), and (\ref{pi2d}), only the diagrams proportional to the coefficient $g^{(6)\mu\nu\alpha\beta\gamma}$ contribute. In contrast, for the coefficient $if^{(6)\alpha\beta\gamma}$, all contributions vanish after computing the traces, and the coefficient $e^{(6)\alpha\beta\gamma}$ likewise does not contribute to the generation of the CFJ term. The remaining contributions can be denoted as $\Pi^{\gamma\delta}_{2a} \rightarrow \Pi^{\gamma\delta}_{CFJ,2a} = \Pi^{\gamma\delta}_{2a,1} + \Pi^{\gamma\delta}_{2a,2} + \Pi^{\gamma\delta}_{2a,3} + \Pi^{\gamma\delta}_{2a,4} + \Pi^{\gamma\delta}_{2a,5}$, where 
\begin{eqnarray}\label{pi22}
    \Pi_{2a,1}^{\gamma\delta} &=& -e^2\int\textstyle{\frac{d^4p}{(2\pi)^4}} \mathrm{tr}\,S(p)\gamma^{\gamma}S(p)\textstyle{\frac{1}{2}}g^{(6)\mu\nu\alpha\beta\kappa}\sigma_{\mu\nu}(-k)_{\alpha}p_{\beta}p_{\kappa}S(p)\gamma^{\delta},\\
    \Pi_{2a,2}^{\gamma\delta} &=& -e^2\int \textstyle{\frac{d^4p}{(2\pi)^4}} \mathrm{tr}\,S(p)\gamma^{\gamma}S(p)\frac{1}{2}g^{(6)\mu\nu\alpha\beta\kappa}\sigma_{\mu\nu} p_{\alpha}(-k)_{\beta}p_{\kappa}S(p)\gamma^{\delta},\\
    \Pi_{2a,3}^{\gamma\delta} &=& -e^2\int\textstyle{\frac{d^4p}{(2\pi)^4}} \mathrm{tr}\,S(p)\gamma^{\gamma}S(p)\frac{1}{2}g^{(6)\mu\nu\alpha\beta\kappa}\sigma_{\mu\nu} p_{\alpha} p_{\beta}(-k)_{\kappa}S(p)\gamma^{\delta},\\
    \Pi_{2a,4}^{\gamma\delta} &=& -e^2\int \textstyle{\frac{d^4p}{(2\pi)^4}} \mathrm{tr}\,S(p)\gamma^{\gamma}S(p)\slashed{k}S(p)\textstyle{\frac{1}{2}}g^{(6)\mu\nu\alpha\beta\kappa}\sigma_{\mu\nu} p_{\alpha}p_{\beta}p_{\kappa}S(p)\gamma^{\delta},\\
    \Pi_{2a,5}^{\gamma\delta} &=& -e^2\int \textstyle{\frac{d^4p}{(2\pi)^4}}\mathrm{tr}\,S(p)\gamma^{\gamma}S(p)\textstyle{\frac{1}{2}}g^{(6)\mu\nu\alpha\beta\kappa}\sigma_{\mu\nu} p_{\alpha}p_{\beta}p_{\kappa}S(p)\slashed{k}S(p)\gamma^{\delta},
\end{eqnarray}
as well as $\Pi^{\gamma\delta}_{2b}\rightarrow\Pi^{\gamma\delta}_{CFJ,2b}$, with
\begin{eqnarray}
    \Pi_{CFJ,2b}^{\gamma\delta} &=& -e^2\int \textstyle{\frac{d^4p}{(2\pi)^4}} \mathrm{tr}\,S(p)\textstyle{\frac{1}{2}}g^{(6)\mu\nu\alpha\beta\kappa}\sigma_{\mu\nu} p_{\alpha}p_{\beta}p_{\kappa}S(p)\gamma^{\gamma}S(p)\slashed{k}S(p)\gamma^{\delta},
\end{eqnarray}
and $\Pi^{\gamma\delta}_{2c}\rightarrow\Pi^{\gamma\delta}_{CFJ,2c} =\Pi^{\gamma\delta}_{2c,1}+\Pi^{\gamma\delta}_{2c,2}$, with
\begin{eqnarray}
    \Pi_{2c,1}^{\gamma\delta} &=& e^2\int \textstyle{\frac{d^4p}{(2\pi)^4}}\mathrm{tr}\,S(p)\textstyle{\frac{1}{2}}g^{(6)\mu\nu\alpha\beta\gamma}\sigma_{\mu\nu}
   3k_{\alpha}p_{\beta}S(p)\gamma^{\delta},\\
    \Pi_{2c,2}^{\gamma\delta} &=& e^2\int  \textstyle{\frac{d^4p}{(2\pi)^4}}\mathrm{tr}\,S(p)\frac{1}{2}g^{(6)\mu\nu\alpha\beta\gamma}\sigma_{\mu\nu}
    \nabla_{\alpha\beta}(p,0)S(p)\slashed{k}S(p)\gamma^{\delta},
\end{eqnarray}
and finally, $\Pi^{\gamma\delta}_{2d}\rightarrow\Pi^{\gamma\delta}_{CFJ,2d} = \Pi^{\gamma\delta}_{2d,1}+\Pi^{\gamma\delta}_{2d,2}$, where
\begin{eqnarray}
    \Pi_{2d,1}^{\gamma\delta} &=& e^2\int \textstyle{\frac{d^4p}{(2\pi)^4}} \mathrm{tr}\,S(p)\gamma^{\gamma}S(p)\frac{1}{2}g^{(6)\mu\nu\alpha\beta\delta}\sigma_{\mu\nu}
    \left(-3k_{\alpha}p_{\beta} - 3p_{\alpha}k_{\beta} - 3k_{\alpha}p_{\beta} \right),\\
    \Pi_{2d,2}^{\gamma\delta} &=& e^2\int  \textstyle{\frac{d^4p}{(2\pi)^4}} \mathrm{tr}\,S(p)\gamma^{\gamma}S(p)\slashed{k}S(p)\frac{1}{2}g^{(6)\mu\nu\alpha\beta\delta}\sigma_{\mu\nu}
    \nabla_{\alpha\beta}(p,0).
\end{eqnarray}

Next, we examine the contributions arising from~(\ref{pi2e}) and (\ref{pi2f}), so that $\Pi^{\gamma\delta}_{2e}\rightarrow\Pi^{\gamma\delta}_{CFJ,2e}$ and $\Pi^{\gamma\delta}_{2f}\rightarrow\Pi^{\gamma\delta}_{CFJ,2f}$, which are given by
\begin{eqnarray}
     \label{CFJ,2e}\Pi^{\gamma\delta}_{CFJ,2e} &=& ie^2\int  \textstyle{\frac{d^4p}{(2\pi)^4}} \mathrm{tr}\,S(p)\gamma^{\gamma}S(p)e_{F}^{(6)\alpha\beta\delta}p_{\alpha}(-k)_{\beta},\\
     \label{CFJ,2f}\Pi^{\gamma\delta}_{CFJ,2f}&=& ie^2\int  \textstyle{\frac{d^4p}{(2\pi)^4}} \mathrm{tr}\,S(p)e_{F}^{(6)\alpha\beta\gamma}p_{\alpha}k_{\beta}S(p)\gamma^{\delta},
\end{eqnarray}
where we have used the fact that the terms involving $if_{F}^{(6)\alpha\beta\gamma}$ and $g_{F}^{(6)\mu\nu\alpha\beta\gamma}$ vanish when taking the trace. Note that there are no contributions from the amplitudes in Eqs.~(\ref{pi2g}) and~(\ref{pi2h}), since they contain at least two derivatives. At this stage, we are interested only in single-derivative terms. Consequently, we set $\Pi^{\gamma\delta}_{2g}\to\Pi^{\gamma\delta}_{CFJ,2g}=0$ and $\Pi^{\gamma\delta}_{2h}\to\Pi^{\gamma\delta}_{CFJ,2h}=0$.

We now present the results for the individual coefficient contributions arising from each of the above expressions. In this context, after calculating the trace, the total contribution to the coefficient $g^{(6)\mu\nu\alpha\delta\gamma}$ can be written as 
\begin{eqnarray}\label{CFJg0}
    \Pi^{\gamma\delta}_{CFJ,g} &=& \textstyle{\frac{1}{2}}\Pi^{\gamma\delta}_{CFJ,2a} + \textstyle{\frac{1}{2}}\Pi^{\gamma\delta}_{CFJ,2b} + \textstyle{\frac{1}{2}}\Pi^{\gamma\delta}_{CFJ,2c} + \textstyle{\frac{1}{2}}\Pi^{\gamma\delta}_{CFJ,2d},
\end{eqnarray}
where 
\begin{eqnarray}
\Pi^{\gamma\delta}_{CFJ,2a}&=&-\frac{2ie^2m}{D}\mu^{4-D}\int \frac{d^Dp}{(2\pi)^D} \frac{p^2}{(p^2-m^2)^{3}}g^{(6)\mu\nu\alpha\beta\kappa}[(m^2-p^2)(\delta^{\gamma}_{\ \nu} \delta^{\delta}_{\ \mu}-\delta^{\gamma}_{\ \mu } \delta^{\delta}_{\ \nu })\nonumber\\
&&\times(k_{\kappa } g_{\alpha  \beta }+k_{\beta } g_{\alpha  \kappa }+k_{\alpha } g_{\beta  \kappa })]\nonumber\\
&&+\frac{4ie^2m}{D(D+2)}\mu^{4-D}\int \frac{d^Dp}{(2\pi)^D}\frac{p^4}{(p^2-m^2)^{4}}g^{(6)\mu\nu\alpha\beta\kappa}[(m^2-p^2)(2 k_{\alpha } g_{\beta  \kappa } (\delta^{\gamma}_{\ \nu } \delta^{\delta}_{\ \mu }-\delta^{\gamma}_{\ \mu } \delta^{\delta}_{\ \nu })\nonumber\\
&&+2 k_{\kappa } g_{\alpha  \beta } \delta^{\gamma}_{\ \nu } \delta^{\delta}_{\ \mu }-2 k_{\kappa } g_{\alpha  \beta } \delta^{\gamma}_{\ \mu } \delta^{\delta}_{\ \nu }-k_{\mu } \delta^{\delta}_{\ \alpha} g_{\beta  \kappa } \delta^{\gamma}_{\ \nu}-k_{\mu } g_{\alpha  \beta } \delta^{\gamma}_{\ \nu } \delta^{\delta}_{\ \kappa }+k_{\mu } \delta^{\gamma}_{\ \alpha} g_{\beta\kappa } \delta^{\delta}_{\ \nu }\nonumber\\
&&+k_{\mu } g_{\alpha  \beta } \delta^{\gamma}_{\ \kappa } \delta^{\delta}_{\nu }+k_{\nu } (\delta^{\gamma}_{\ \mu } (g_{\alpha  \beta } \delta^{\delta}_{\ \kappa }+\delta^{\delta}_{\ \alpha} g_{\beta  \kappa })-\delta^{\delta}_{\ \mu } (g_{\alpha  \beta } \delta^{\gamma}_{\ \kappa }+\delta^{\gamma}_{\ \alpha} g_{\beta  \kappa }))\nonumber\\
&&+ g_{\alpha  \kappa } (k_{\nu } (\delta^{\delta}_{\ \beta} \delta^{\gamma}_{\ \mu }-\delta^{\gamma}_{\ \beta} \delta^{\delta}_{\ \mu })+k_{\mu } (\delta^{\gamma}_{\ \beta} \delta^{\delta}_{\ \nu }-\delta^{\delta}_{\ \beta} \delta^{\gamma}_{\ \nu })+2 k_{\beta } (\delta^{\gamma}_{\ \nu } \delta^{\delta}_{\ \mu }-\delta^{\gamma}_{\ \mu } \delta^{\delta}_{\ \nu })))],\\
\Pi^{\gamma\delta}_{CFJ,2b}&=&-\frac{4ie^2m}{D(D+2)}\mu^{4-D}\int \frac{d^Dp}{(2\pi)^D} \frac{p^4}{(p^2-m^2)^{4}}g^{(6)\mu\nu\alpha\beta\kappa}[(m^2-p^2)(k_{\mu } (\delta^{\gamma}_{\ \nu } (g_{\alpha\beta } \delta^{\delta}_{\ \kappa }\nonumber\\
&&+\delta^{\delta}_{\ \alpha} g_{\beta  \kappa }+g_{\alpha  \kappa } \delta^{\delta}_{\ \beta})-\delta^{\delta}_{\ \nu } (g_{\alpha\beta } \delta^{\gamma}_{\ \kappa }+\delta^{\gamma}_{\ \alpha}g_{\beta  \kappa }+g_{\alpha  \kappa } \delta^{\gamma }_{\ \beta}))+k_{\nu } (\delta^{\delta}_{\ \mu } (g_{\alpha  \beta } \delta^{\gamma}_{\ \kappa}+g^{\gamma}_{\ \alpha} g_{\beta  \kappa }\nonumber\\
&&+g_{\alpha  \kappa } \delta^{\gamma}_{\ \beta})-\delta^{\gamma}_{\ \mu } (g_{\alpha\beta } \delta^{\delta}_{\ \kappa}+\delta^{\delta}_{\ \alpha} g_{\beta  \kappa }+g_{\alpha\kappa } \delta^{\delta}_{\ \beta})))],\\
\Pi^{\gamma\delta}_{CFJ,2c}&=&\frac{6ie^2}{D}\mu^{4-D}\int \frac{d^Dp}{(2\pi)^D} \frac{p^2}{(p^2-m^2)^{3}}g^{(6)\mu\nu\alpha\beta\delta}[m^3 k_{\mu } \delta_{\alpha  \beta } \delta^{\gamma}_{\ \nu } - m^3 k_{\mu } g_{\alpha  \beta } \delta^{\gamma}_{\ \nu}\nonumber\\
&&-mp^2 k_{\mu } g_{\alpha\beta } \delta^{\gamma}_{\ \nu }+mp^2 k_{\nu } g_{\alpha  \beta } \delta^{\gamma}_{\ \mu }],\\
\Pi^{\gamma\delta}_{CFJ,2d}&=&\frac{6ie^2}{D}\mu^{4-D}\int \frac{d^Dp}{(2\pi)^D} \frac{p^2}{(p^2-m^2)^{3}}g^{(6)\mu\nu\alpha\beta\gamma}(-m^3  k_{\mu} g_{\alpha\beta} \delta^{\delta}_{\ \nu}+m^3 k_{\nu } g_{\alpha\beta } \delta^{\delta}_{\ \mu }\nonumber\\
&&+mp^2k_{\mu } g_{\alpha  \beta }\delta^{\delta}_{\ \nu }-mp^2 k_{\nu } g_{\alpha\beta}  \delta^{\delta}_{\ \mu }).
\end{eqnarray}
Each of these contributions is calculated separately using dimensional regularization. Accordingly, we obtain the following results:
\begin{eqnarray}
\Pi^{\gamma\delta}_{CFJ,2a}&=&g^{(6)\mu\nu\alpha\beta\kappa}\frac{ 2^{-D-1} \pi ^{1-\frac{D}{2}} e^2 \mu ^{4-D} m^{D-1} ( \csc(\pi  D)/2 )}{\Gamma (\frac{D}{2})}\nonumber\\
&&\times(k_{\mu } (\delta^{\gamma}_{\ \nu } (g_{\alpha\beta } \delta^{\delta}_{\ \kappa }+\delta^{\delta}_{\ \alpha} g_{\beta  \kappa }+g_{\alpha\kappa } \delta^{\delta}_{\ \beta})-\delta^{\delta}_{\ \nu } (g_{\alpha  \beta } \delta^{\gamma}_{\ \kappa }+\delta^{\gamma}_{\ \alpha} g_{\beta\kappa }+g_{\alpha \kappa } g^{\gamma}_{\ \beta}))\nonumber\\
&&+k_{\nu } (\delta^{\delta}_{\ \mu } (g_{\alpha  \beta } \delta^{\gamma}_{\ \kappa }+\delta^{\gamma}_{\ \alpha} g_{\beta  \kappa }+g_{\alpha  \kappa } \delta^{\gamma}_{\ \beta})-\delta^{\gamma}_{\ \mu } (g_{\alpha\beta } \delta^{\delta}_{\ \kappa }+\delta^{\delta}_{\ \alpha} g_{\beta  \kappa }+g_{\alpha  \kappa } \delta^{\delta}_{\ \beta}))),
\end{eqnarray}
 \begin{eqnarray}
\Pi^{\gamma\delta}_{CFJ,2b} &=& g^{(6)\mu\nu\alpha\beta\kappa}\frac{ 2^{-D} \pi ^{1-\frac{D}{2}} e^2 \mu ^{4-D} m^{D-1} ( \csc (\pi  D)/2)}{\Gamma (\frac{D}{2})}\nonumber\\
&&\times(k_{\mu } (\delta^{\gamma}_{\ \nu } (g_{\alpha  \beta } \delta^{\delta}_{\ \kappa }+\delta^{\delta}_{\ \alpha} g_{\beta  \kappa }+g_{\alpha \kappa } \delta^{\delta}_{\ \beta})-\delta^{\delta}_{\ \nu } (g_{\alpha  \beta } \delta^{\gamma}_{\ \kappa }+\delta^{\gamma}_{\ \alpha} g_{\beta  \kappa }+g_{\alpha  \kappa } \delta^{\gamma}_{\ \beta}))\nonumber\\
&&+k_{\nu } (\delta^{\delta}_{\ \mu } (g_{\alpha  \beta } \delta^{\gamma}_{\ \kappa }+\delta^{\gamma}_{\ \alpha} g_{\beta  \kappa }+g_{\alpha  \kappa } \delta^{\gamma}_{\ \beta})-\delta^{\gamma}_{\mu } (g_{\alpha\beta } \delta^{\delta}_{\ \kappa }+\delta^{\delta}_{\ \alpha} g_{\beta  \kappa }+g_{\alpha  \kappa } \delta^{\delta}_{\ \beta}))),
\end{eqnarray}
 \begin{eqnarray}
\Pi^{\gamma\delta}_{CFJ,2c} &=& - g^{(6)\mu\nu\alpha\beta\delta}\frac{3\ 2^{-D} \pi ^{1-\frac{D}{2}} e^2 \mu ^{4-D} m^{D-1}  \csc (\pi  D)}{ \Gamma \left(\frac{D}{2}\right)}g_{\alpha  \beta }(k_{\mu } \delta^{\gamma}_{\ \nu }-k_{\nu } \delta^{\gamma}_{\ \mu }),
 \end{eqnarray}
 \begin{eqnarray}
\Pi^{\gamma\delta}_{CFJ,2d} &=& g^{(6)\mu\nu\alpha\beta\gamma}  \frac{3\ 2^{-D} \pi ^{1-\frac{D}{2}} e^2 \mu ^{4-D} m^{D-1}  \csc (\pi  D)}{ \Gamma \left(\frac{D}{2}\right)}g_{\alpha  \beta }  (k_{\mu } \delta^{\delta}_{\ \nu }-k_{\nu } \delta^{\delta}_{\ \mu }).
 \end{eqnarray}

At this point, we once more carry out a series expansion and, using the specific structure of the operator, the coefficient $g^{(6)\mu\nu\rho\alpha\beta}$ can be written as
\begin{equation}\label{coeficiente_g}
    g^{(6)\mu\nu\rho\alpha\beta} = \frac{1}{3}g^{\alpha\beta}
   \epsilon^{\rho\mu\nu\lambda}g_{\lambda}+\frac{1}{3}g^{\alpha\rho}
   \epsilon^{\beta\mu\nu\lambda}g_{\lambda}+ \frac{1}{3}g^{\beta\rho}
   \epsilon^{\alpha\mu\nu\lambda}g_{\lambda},
\end{equation}
where this tensor has been defined so that it is antisymmetric in the first pair of indices and symmetric in the last three ones. Therefore, upon adopting the above decomposition, the polarization amplitude (\ref{CFJg0}) gives a vanishing contribution, namely,
\begin{eqnarray}\label{CFJg}
    \Pi^{\gamma\delta}_{CFJ,g} =  0.
\end{eqnarray}

For the polarization amplitudes associated with the coefficient $e_{F}^{(6)\alpha\beta\gamma}$, once the trace is evaluated, the contributions~\eqref{CFJ,2e} and~\eqref{CFJ,2f} yield
\begin{subequations}
    \begin{eqnarray}
    \Pi^{\gamma\delta}_{CFJ,2e} &=& -\frac{8ie^2m}{D}\mu^{4-D}\int\frac{d^Dp}{(2\pi)^D}\frac{p^2}{(p^2 -m^2)^2}\delta^{\gamma}_{\ \alpha}\delta^{\delta}_{\ \lambda}k_{\beta}e_F^{(6)\alpha\beta\lambda},\\
     \Pi^{\gamma\delta}_{CFJ,2f} &=& \frac{8ie^2m}{D}\mu^{4-D}\int\frac{d^Dp}{(2\pi)^D}\frac{p^2}{(p^2 -m^2)^2}\delta^{\gamma}_{\ \kappa}\delta^{\delta}_{\ \beta}k_{\beta}e_F^{(6)\alpha\beta\kappa}.
\end{eqnarray}
\end{subequations}
By explicitly carrying out the integrations, we find that the amplitudes corresponding to the coefficient $e_F^{(6)\alpha\beta\gamma}$ are 
\begin{subequations}
    \begin{eqnarray}
    \Pi^{\gamma\delta}_{CFJ,2e} &=&  \frac{2^{2-D} \pi ^{1-\frac{D}{2}} e^2 \mu ^{4-D} m^{D-1}  \csc (\pi  D) }{ \Gamma \left(\frac{D}{2}\right)}\epsilon^{\gamma\delta\alpha\beta}k_{\alpha}(e_F)_{\beta},\\
    \Pi^{\gamma\delta}_{CFJ,2f} &=&  \frac{2^{2-D} \pi ^{1-\frac{D}{2}} e^2 \mu ^{4-D} m^{D-1}  \csc (\pi  D) }{\Gamma \left(\frac{D}{2}\right)}\epsilon^{\gamma\delta\alpha\beta}k_{\alpha}(e_F)_{\beta},
    \end{eqnarray}
\end{subequations}
where we have used the series expansion of the coefficient $e_{F}^{(6)\alpha\beta\gamma}$, as presented in Eq.~\eqref{coeficiente_eFm_DF}. Thus, the total result can be expressed as 
\begin{eqnarray}\label{CFJeF}
    \Pi^{\gamma\delta}_{CFJ,e_F} &=& \Pi^{\gamma\delta}_{CFJ,1b} +\textstyle{\frac{1}{2}}\Pi^{\gamma\delta}_{CFJ,2e}+ \textstyle{\frac{1}{2}} \Pi^{\gamma\delta}_{CFJ,2f}\nonumber\\
    &=& - \frac{e^2m^3}{\pi ^2 \epsilon^{\prime}} \epsilon^{\gamma\delta\alpha\beta}k_{\alpha}(e_F)_{\beta} -\frac{e^2m^3}{2\pi^2}\epsilon^{\gamma\delta\alpha\beta}k_{\alpha}(e_F)_{\beta}.
\end{eqnarray}

Note that from the results obtained in \eqref{CFJg}, \eqref{CFJeF}, and taking into account that $\Pi^{\gamma\delta}_{CFJ,1c}\rightarrow\Pi^{\gamma\delta}_{CFJ,m_{DF}}$ in \eqref{CFJ1c}, where $(m_{DF})_{\beta}$ is the negative dimension (explicitly, its mass dimension is $-2$) axial vector, whose presence can naturally allow for arising the higher-derivative CFJ-like correction, we can write an equation for the CFJ term related to these coefficients as follows: 
\begin{eqnarray}\label{PiCFJ}
\Pi^{\gamma\delta}_{CFJ} &=& \Pi^{\gamma\delta}_{CFJ,g} + \Pi^{\gamma\delta}_{CFJ,e_F} + \Pi^{\gamma\delta}_{CFJ, m_{DF}} \nonumber\\
&=&-\frac{e^2m^3}{2\pi^{2}}\epsilon^{\gamma\delta\alpha\beta}k_{\alpha}(2(e_{F})_{\beta}+i(m_{DF})_{\beta})\left(\frac{1}{\epsilon^{\prime}}+\frac{1}{2}\right).
\end{eqnarray}
We note that the CFJ term turns out to diverge, unlike almost all generating schemes where it is finite (see, e.g. \cite{Colladay:1998fq,Jackiw:1999qq,Andrianov:2001zj,Chung:1998jv,Chung:1999pt,Mariz:2005jh} and many other papers), unless we apply a special relation between LV parameters; on the other hand, it can mean that in this case, if the divergence survives, there is no possibility to generate a chiral anomaly within this scheme, while usually the ambiguity of the finite result for the CFJ term is related with this anomaly. At the same time, if we impose the special relation between the LV parameters, that is, $2(e_F)_\beta +i (m_{DF})_{\beta}=0$, divergent and finite parts of the self-energy tensor are ruled out. Therefore, we can conclude that in this case, the vanishing of the divergence implies the vanishing of the finite part of the CFJ term. Nevertheless, it is interesting to note that the zero result for the CFJ term is advantageous in a certain sense, since it allows for the gauge invariance not only of the effective action, but also of the effective Lagrangian itself; see the discussion in \cite{Altschul:2019eip}. Moreover, it is important to note that zero results for the CFJ term are known to arise for various LV extensions of QED; see e.g. \cite{Coleman:1998ti,dim51}.

In the next section, we will perform the calculation of these contributions using the Feynman parametrization method. This procedure allows the rearrangement of the denominators of momentum integrals, enabling the systematic derivation of all orders in derivatives in the development of the effective action.

\section{Feynman Parametrization}

We will employ dimensional regularization to evaluate the integrals. However, unlike the previous case, exact propagators will be considered. Thus, after computing the trace and introducing the Feynman parameter $x$, the expressions with denominators depending on the external momentum $k$ will be evaluated.

Within this section, to implement the Feynman parametrization method, we will use the relation
\begin{eqnarray}
    \frac{1}{AB} = \int^{1}_{0} \frac{dx}{\left(Ax + B(1-x)\right)^2},
\end{eqnarray}
which makes it possible to explicitly calculate each contribution corresponding to the terms presented in Eqs.~\eqref{pi1} and \eqref{pi2a}--\eqref{pi2h}. Thus, when considering the coefficients $i f^{(6)\alpha\beta\gamma}\gamma_{5}$ and $i f_{F}^{(6)\alpha\beta\gamma}\gamma_{5}$, we verify that their respective contributions to the trace calculations vanish. Therefore, it is valid to write
\begin{equation}
{\cal L}_{f} = \left(i \Pi_{1a,f}^{\gamma\delta} +\frac{i}{2}\Pi_{2a,f}^{\gamma\delta} + \frac{i}{2}\Pi_{2b,f}^{\gamma\delta} + \frac{i}{2}\Pi_{2c,f}^{\gamma\delta} + \frac{i}{2}\Pi_{2d,f}^{\gamma\delta}\right)A_{\gamma}A_{\delta}= 0
\end{equation}
and
\begin{equation}
{\cal L}_{f_F} = \left(i \Pi_{1b,f_{F}}^{\gamma\delta} + \frac{i}{2}\Pi_{2e,f_{F}}^{\gamma\delta} + \frac{i}{2}\Pi_{2f,f_{F}}^{\gamma\delta}\right)A_{\gamma}A_{\delta}= 0.
\end{equation}

It is also easy to see that, when considering the coefficient $m_{5DF}^{(6)\alpha\beta\gamma}$, the trace calculation also yields null results. In particular, the terms proportional to ${\rm tr}(\slashed{p}+m)\gamma_{5} = 0$ vanish identically. Furthermore, we also verify that ${\rm tr}(\slashed{p}+m)\gamma^{\delta}(\slashed{p} - \slashed{k}+m) \gamma_{5} = 0$ and ${\rm tr}(\slashed{p}+m)\gamma_{5}(\slashed{p}- \slashed{k}+m)\gamma^{\delta}=0$. Then, we have 
\begin{equation}
{\cal L}_{m_{5DF}} = \left(i \Pi_{1c,m_{5DF}}^{\gamma\delta} + \frac{i}{2}\Pi_{2g,m_{5DF}}^{\gamma\delta} + \frac{i}{2}\Pi_{2h,m_{5DF}}^{\gamma\delta}\right)A_{\gamma}A_{\delta}= 0.
\end{equation}

Following the analysis, after performing the trace and introducing the Feynman parameter $x$, we obtain the resulting expressions for the terms proportional to $e_{F}^{(6)\alpha\beta\gamma}$:
\begin{eqnarray}
\Pi_{2e,e_{F}}^{\gamma\delta} = 4 i e^2 m \, \mu^{4-D}\int^{1}_{0}dx\int\frac{d^Dp}{(2\pi)^D}\frac{k_{\beta } q_{\alpha } g^{\delta}_{\ \lambda }}{(p^2-M^2)^2} (k^{\gamma }-2 q^{\gamma })e_{F}^{(6)\alpha\beta\lambda},
\end{eqnarray}
\begin{eqnarray}
\Pi_{2f,e_{F}}^{\gamma\delta} = 4 i e^2 m \, \mu^{4-D}\int^{1}_{0}dx\int\frac{d^Dp}{(2\pi)^D}\frac{k_{\beta } g^{\gamma}_{\ \kappa }}{(p^2-M^2)^2} (k_{\alpha }-q_{\alpha }) (k^{\delta }-2 q^{\delta })e_{F}^{(6)\alpha\beta\kappa},
\end{eqnarray}
where we define $q_{\mu} = p_{\mu} + (1-x)k_{\mu}$ as the shifted momentum and write the mass as $M^2 = m^2 + x(x-1)k^2$. Then, after evaluating the integrals, the divergent terms can be written explicitly as
\begin{eqnarray}
\Pi_{2e,e_{F}}^{\gamma\delta}+\Pi_{2f,e_{F}}^{\gamma\delta}&=&e_{F}^{(6)\kappa\beta\lambda}\frac{e^2 m^3}{2 \pi ^2 \epsilon^{\prime}}(k_{\beta } \delta^{\gamma}_{\ \kappa } \delta^{\delta}_{\ \lambda })+ e_{F}^{(6)\alpha\beta\lambda}\frac{e^2 m}{12 \pi ^2 \epsilon^{\prime}}(k_{\beta } \delta^{\delta}_{\ \lambda }((6 m^2-k^2) \delta^{\gamma }_{\ \alpha}+k_{\alpha } k^{\gamma }))\nonumber\\
&&-e_{F}^{(6)\alpha\beta\kappa} \frac{e^2 m }{12 \pi ^2 \epsilon^{\prime}}(k_{\beta } \delta^{\gamma}_{\ \kappa } ((6 m^2-k^2) {\delta^\delta}_\alpha+k_{\alpha } k^{\delta })) +\mathrm{finite\ terms}. 
\end{eqnarray}
These divergent terms contribute directly to the effective Lagrangian for the gauge field. By combining the relevant amplitudes, the corresponding term in the effective action can be expressed as follows:
\begin{eqnarray}\label{LeF}
{\cal L}_{e_F} &=& \left( i\Pi_{1b, e_{F}}^{\gamma\delta} + \frac{i}{2}\Pi_{2e,e_{F}}^{\gamma\delta}+  \frac{i}{2}\Pi_{2f,e_{F}}^{\gamma\delta}\right)A_{\gamma}A_{\delta}\nonumber\\
&=& -\frac{ie^2 \left(12 m^3-k^2 m\right)}{12 \pi ^2 \epsilon^{\prime}} \epsilon ^{\gamma  \delta \alpha\beta }k_{\alpha}(e_{F})_{\beta}A_{\gamma}A_{\delta} + \mathrm{finite\ terms},
\end{eqnarray}
where we require the coefficient $e_{F}^{\alpha\beta\gamma}$ to be antisymmetric in all indices, as stated in Eq.~\eqref{eF}. This condition is essential, as it determines which tensor combinations effectively contribute to the divergent terms in the effective action.

Similarly, for the coefficient $m_{DF}^{(6)\delta\beta\gamma}$, after performing the trace and introducing the Feynman parameter $x$, we obtain
\begin{eqnarray}
\Pi_{2g,m_{DF}}^{\gamma\delta} = 4 e^2 m \,\mu^{4-D}\int^{1}_{0}dx\int\frac{d^Dp}{(2\pi)^D}\frac{k_{\alpha } k_{\beta } g^{\delta}_{\ \lambda }}{(p^2-M^2)^2}(k^{\gamma }-2 q^{\gamma })m_{DF}^{(6)\alpha\beta\lambda},
\end{eqnarray}
\begin{eqnarray}
\Pi_{2h,m_{DF}}^{\gamma\delta} = 4 e^2 m \,\mu^{4-D}\int^{1}_{0}dx\int\frac{d^Dp}{(2\pi)^D}\frac{k_{\alpha } k_{\beta } g^{\gamma}_{\ \lambda }}{(p^2-M^2)^2}(k^{\delta }-2 q^{\delta })m_{DF}^{(6)\alpha\beta\lambda}.
\end{eqnarray}
Then, after performing the integrals, we obtain
\begin{eqnarray}
\Pi_{2g,m_{DF}}^{\gamma\delta}+\Pi_{2h,m_{DF}}^{\gamma\delta} &=& 0.
\end{eqnarray}
Thus, only the amplitude $\Pi_{1c,m_{DF}}^{\gamma\delta}$ contributes to the Lagrangian. Consequently, using the decomposition for this coefficient (\ref{mDF}), the corresponding divergent term in the effective Lagrangian can be written as:
\begin{eqnarray}\label{LmDF}
{\cal L}_{m_{DF}} &=& \left( i \Pi_{1c,m_{DF}}^{\gamma\delta} + \frac{i}{2}\Pi_{2g,m_{DF}}^{\gamma\delta}  + \frac{i}{2}\Pi_{2h,m_{DF}}^{\gamma\delta}\right)A_{\gamma}A_{\delta}\nonumber\\
&=& \frac{e^2 m^3}{2 \pi ^2 \epsilon^{\prime}}\epsilon^{\gamma\delta\alpha\beta}k_{\alpha}
     (m_{DF})_{\beta}A_{\gamma}A_{\delta} +\mathrm{finite\ terms}.
\end{eqnarray}

For the divergent terms associated with the coefficient $g^{(6)\mu\nu\alpha\beta\gamma}$, it is necessary to sum the following  contributions:
\begin{eqnarray}
\Pi_{2a,g}^{\gamma\delta} &=& 2ime^2 \mu^{4-D} \int^{1}_{0}dx(2x-2)\int\frac{d^Dp}{(2\pi)^D}\frac{(k_{\alpha }-q_{\alpha })(k_{\beta }-q_{\beta })(k_{\kappa }-q_{\kappa })}{(p^2-M^2)^3}g^{(6)\mu\nu\alpha\beta\kappa}\nonumber\\
&&\times (2 (k^{\gamma }-q^{\gamma })(k_{\mu } \delta^{\delta}_{\ \nu }-k_{\nu } \delta^{\delta}_{\ \mu })+\delta^{\gamma}_{\ \mu } (-(\delta^{\delta}_{\ \nu }(k^2+m^2-q^2)+2 k_{\nu } (q^{\delta }-k^{\delta })))\nonumber\\
&&+g^{\gamma}_{\ \nu } (\delta^{\delta}_{\ \mu } (k^2+m^2-q^2)+2 k_{\mu }(q^{\delta }-k^{\delta }))),
\end{eqnarray}
\begin{eqnarray}
\Pi_{2b,g}^{\gamma\delta} &=& 2ime^2 \mu^{4-D} \int^{1}_{0}dx(2x)\int\frac{d^Dp}{(2\pi)^D}\frac{q_{\alpha } q_{\beta } q_{\kappa }}{(p^2-M^2)^3}g^{(6)\mu\nu\alpha\beta\kappa}\nonumber\\
&&\times (\delta^{\gamma}_{\ \mu } (\delta^{\delta}_{\ \nu} (2 k\cdot q+m^2-q^2)-2 k_{\nu } q^{\delta })+\delta^{\gamma}_{\ \nu } (\delta^{\delta}_{\ \mu } (-2 k\cdot q-m^2+q^2)+2 k_{\mu } q^{\delta })\nonumber\\
&&+2 q^{\gamma } (k_{\nu } \delta^{\delta}_{\ \mu }-k_{\mu } \delta^{\delta}_{\ \nu })),
\end{eqnarray}
\begin{eqnarray}
\Pi_{2c,g}^{\gamma\delta} &=& -2ime^2 \mu^{4-D} \int^{1}_{0}dx\int\frac{d^Dp}{(2\pi)^D}\frac{g^{\,\,\delta}_{\lambda}}{(p^2-M^2)^2}g^{(6)\mu\nu\alpha\beta\gamma}\nonumber\\
&&\times (3 q_{\alpha } (q_{\beta }-k_{\beta })+k_{\alpha } (7 k_{\beta }-6 q_{\beta }))(k_{\mu } \delta^{\gamma}_{\ \nu }-k_{\nu } \delta^{\gamma}_{\ \mu }),
\end{eqnarray}
\begin{eqnarray}
\Pi_{2d,g}^{\gamma\delta} &=& 2ime^2 \mu^{4-D} \int^{1}_{0}dx\int\frac{d^Dp}{(2\pi)^D}\frac{g^{\,\,\gamma}_{\lambda}}{(p^2-M^2)^2}g^{(6)\mu\nu\alpha\beta\delta}\nonumber\\
&&\times (3 q_{\beta } (k_{\alpha }+q_{\alpha })+k_{\alpha } k_{\beta }) (k_{\mu } \delta^{\delta}_{\ \nu }-k_{\nu } \delta^{\delta}_{\ \mu }).
\end{eqnarray}
Thus, by consistently performing the integrations and carrying out the contractions according to the proposed structure of the coefficient $g^{(6)\mu\nu\alpha\beta\gamma}$ (\ref{coeficiente_g}), the divergent part can be expressed as
\begin{eqnarray}
\Pi_{2a,g}^{\gamma\delta}+\Pi_{2b,g}^{\gamma\delta}+\Pi_{2c,g}^{\gamma\delta}+\Pi_{2d,g}^{\gamma\delta} &=& \frac{e^2k^2m}{2\pi^2\epsilon^{\prime}}\epsilon^{\gamma\delta\alpha\beta}k_{\alpha}g_{\beta} + \mathrm{finite\ terms}.
\end{eqnarray}
Therefore, keeping only the divergent terms, our contribution to the effective Lagrangian reads:
\begin{eqnarray}\label{Lg}
{\cal L}_{g} &=& \left(i \Pi_{1a,g}^{\gamma\delta} +\frac{i}{2}\Pi_{2a,g}^{\gamma\delta} + \frac{i}{2}\Pi_{2b,g}^{\gamma\delta} + \frac{i}{2}\Pi_{2c,g}^{\gamma\delta} + \frac{i}{2}\Pi_{2d,g}^{\gamma\delta}\right)A_{\gamma}A_{\delta}\nonumber\\
&=& \frac{ie^2k^2m}{4\pi^2\epsilon^{\prime}}\epsilon^{\gamma\delta\alpha\beta}k_{\alpha}g_{\beta}+ \mathrm{finite\ terms}.
\end{eqnarray}

The integrals associated with the coefficient $g_{F}^{(6)\mu\nu\alpha\beta\gamma}$ are written as
\begin{eqnarray}
\Pi^{\gamma\delta}_{2g,g_{F}}&=&2e^2m\mu^{4-D}\int^{1}_{0}dx\int\frac{d^{D}p}{(2\pi)^{D}}\frac{k_{\beta } q_{\alpha } \delta^{\delta}_{\ \lambda} \left(k_{\nu } \delta^{\gamma}_{\ \mu}-k_{\mu } \delta^{\gamma}_{\ \nu }\right)}{(p^2-M^2)^{2}}\ g_{F}^{(6)\mu\nu\alpha\beta\lambda},\\
\Pi^{\gamma\delta}_{2h,g_{F}}&=&2e^2m\mu^{4-D}\int^{1}_{0}dx\int\frac{d^{D}p}{(2\pi)^{D}}\frac{k_{\beta } \delta^{\gamma}_{\ \lambda } \left(k_{\alpha }-q_{\alpha }\right) \left(k_{\mu } \delta^{\delta}_{\ \nu }-k_{\nu } \delta^{\delta}_{\ \mu }\right)}{(p^2-M^2)^{2}}\ g_{F}^{(6)\mu\nu\alpha\beta\lambda},
\end{eqnarray}
so that, upon evaluating the integrals, we obtain
\begin{eqnarray}
\Pi^{\gamma\delta}_{2g,g_{F}}+\Pi^{\gamma\delta}_{2h,g_{F}}&=&-\frac{i e^2 m}{8\pi^{2}\epsilon^{\prime}} k_{\alpha } k_{\beta }(-k_{\nu } \delta^{\gamma}_{\ \mu } \delta^{\delta}_{\ \lambda }+k_{\mu } \delta^{\gamma}_{\ \nu } \delta^{\delta}_{\ \lambda }+\delta^{\gamma}_{\ \lambda } (k_{\nu } \delta^{\delta}_{\ \mu }-k_{\mu } \delta^{\delta}_{\ \nu }))g_{F}^{(6)\mu\nu\alpha\beta\lambda} \nonumber\\
&&+\mathrm{finte\ terms}.
\end{eqnarray}
We now decompose the LV coefficient $g_{F}^{(6)\mu\nu\alpha\beta\gamma}$ such that its contribution is antisymmetric in the index pairs $\mu,\nu$ and $\beta,\gamma$, due to the structure of the operator with which this coefficient is contracted. Thus, we can write
\begin{eqnarray}
g_{F}^{(6)\mu\nu\alpha\beta\gamma} &=& g^{\alpha  \beta } \epsilon ^{\gamma  \mu  \nu  \lambda}(g_F)_\lambda-g^{\alpha  \gamma } \epsilon ^{\beta  \mu  \nu  \lambda}(g_F)_\lambda+g^{\beta  \mu } \epsilon ^{\alpha  \gamma  \nu  \lambda}(g_F)_\lambda-g^{\gamma  \mu } \epsilon ^{\alpha  \beta  \nu  \lambda}(g_F)_\lambda \nonumber\\
&&-g^{\beta  \nu } \epsilon ^{\alpha  \gamma  \mu  \lambda}(g_F)_\lambda+g^{\gamma  \nu } \epsilon ^{\alpha  \beta  \mu  \lambda}(g_F)_\lambda.
\end{eqnarray}
As a consequence of this decomposition, the polarization amplitude associated with this coefficient vanishes, i.e., $\Pi^{\gamma\delta}_{2g,g_{F}}+\Pi^{\gamma\delta}_{2h,g_{F}}=0$, so that we have
\begin{equation}
{\cal L}_{g_F} = \left(i \Pi_{1b,g_F}^{\gamma\delta} + \frac{i}{2}\Pi^{\gamma\delta}_{2g,g_{F}} + \frac{i}{2}\Pi^{\gamma\delta}_{2h,g_{F}}\right)A_{\gamma}A_{\delta}= 0.
\end{equation}

Analogously to the coefficient $g_{F}^{(6)\mu\nu\alpha\gamma}$, we verify that the coefficient $H_{DF}^{\mu\nu\alpha\beta\gamma}$ also leads to trivial contributions after evaluating the corresponding integrals, so that
\begin{eqnarray}
    {\cal L}_{H_{DF}} &=& \left( i\Pi_{1c,H_{DF}}^{\gamma\delta} + \frac{i}{2}\Pi_{2g,H_{DF}}^{\gamma\delta} + \frac{i}{2}\Pi_{2h,H_{DF}}^{\gamma\delta} \right) A_{\gamma}A_{\delta} = 0.
\end{eqnarray}

At this point, we seek to identify relations that allow us to eliminate the divergent terms present in the contributions analyzed, to isolate a finite structure compatible with the higher-derivative CFJ-like term. To this end, we list below all quantities whose expressions contain divergent terms, namely Eqs.~(\ref{LeF}), (\ref{LmDF}), and (\ref{Lg}):
\begin{eqnarray}\label{termos divergentes}
\Pi^{\gamma\delta}_{CFJ} &=& \Pi ^{\gamma\delta}_{e_{F}} + \Pi ^{\gamma\delta}_{m_{DF}} + \Pi ^{\gamma\delta}_{g}\nonumber\\
&=& -\frac{e^2m^3}{2\pi^{2}\epsilon^{\prime}}\epsilon^{\gamma\delta\alpha\beta}k_{\alpha}\bigl(2(e_{F})_{\beta}+i(m_{DF})_{\beta}\bigr) +\frac{e^2k^2m}{12\pi^{2}\epsilon^{\prime}}\epsilon^{\gamma\delta\alpha\beta}k_{\alpha}\bigl((e_{F})_{\beta}+3g_{\beta}\bigr) \nonumber\\
&& +\text{finite terms}.
\end{eqnarray}
We observe that the divergences associated with these structures can be eliminated if certain conditions are satisfied. For example, we can choose
\begin{equation}\label{cond}
(m_{DF})_{\beta}=2i(e_{F})_{\beta}\;\; \text{and}\;\; g_{\beta}=-\frac{1}{3}(e_{F})_{\beta},
\end{equation}
where the first condition cancels the divergent terms proportional to $m^3$, already observed in Eq.~(\ref{PiCFJ}), and the second condition cancels the divergent terms proportional to $k^2 m$, which appear only in the Feynman parametrization.

Then, having found the conditions necessary to eliminate the divergences, we can write the finite terms. Their explicit form is
\begin{eqnarray}
\Pi^{\gamma\delta}_{CFJ} &=&- \frac{e^2m}{36\pi^2\sqrt{k^2(4m^2-k^2)}}\Bigg[ 30m^2\sqrt{k^2(4m^2 - k^2)} - 4(k^2)^{3/2}\sqrt{4m^2-k^2}\nonumber\\
&& -3(k^2-4m^2)^{2}\cot^{-1}\left(\sqrt{\frac{4m^2}{k^2}-1}\right)\Bigg]\epsilon ^{\gamma\delta\alpha\beta}\,(e_{F})_{\beta}k_{\alpha}\nonumber\\
&& -\frac{e^2m}{72\pi^2\sqrt{k^2(4m^2-k^2)}}\Bigg[ 42m^2\sqrt{k^2(4m^2 - k^2)} -17(k^2)^{3/2}\sqrt{4m^2-k^2}\nonumber\\
&& -6(-16k^2m^2+3k^4 + 28m^4)\cot^{-1}\left(\sqrt{\frac{4m^2}{k^2}-1}\right)\Bigg]\epsilon ^{\gamma\delta\alpha\beta}\,g_{\beta}k_{\alpha}\nonumber\\
&& -\frac{ie^2m^3}{4\pi^2}\epsilon^{\gamma\delta\alpha\beta}\,(m_{DF})_{\beta}k_{\alpha},
\end{eqnarray}
or, upon imposing Eq.\ (\ref{cond}), the finite term can be written simply as
\begin{equation}\label{Pifinito}
\Pi^{\gamma\delta}_{CFJ} = -\frac{e^2m}{216\pi^2}\epsilon^{\gamma\delta\alpha\beta}\Bigg[30m^2 - 7k^2 + \frac{(48k^2m^2 - 120m^4)}{\sqrt{k^2(4m^2-k^2)}}\cot^{-1}\left(\sqrt{\frac{4m^2}{k^2}-1}\right)\Bigg]k_{\alpha}(e_{F})_{\beta}.
\end{equation}

At this point, we take the infrared limit, i.e., $k^2 \ll m^2$ ($m \neq 0$), in Eq.~\eqref{Pifinito}, so that the amplitude $\Pi^{\gamma\delta}_{\mathrm{CFJ}}$ can be expressed as
\begin{eqnarray}
\Pi^{\gamma\delta}_{\mathrm{CFJ}} = -\frac{e^2 k^4}{216\pi^2 m}\, \epsilon^{\gamma\delta\alpha\beta}k_{\alpha}(e_{F})_{\beta} + \mathcal{O}\left(\frac{k^6}{m^2}\right).
\end{eqnarray}
We conclude that we have succeeded in generating the five-derivative CFJ-like term, which turns out to be finite, while the three-derivative term, which arises, for example, in schemes employing only minimal couplings \cite{Leite:2013pca}, is absent.

\section{Summary}

We studied the non-minimal, CPT-odd extension of the LV QED involving dimension-6 operators. Within this model, we studied the possibility of arising CFJ-like one-loop quantum corrections. It turns out that in our case the divergent CFJ term arises for certain LV parameters, except for the case of their special relation, which apparently means that for these parameters, there is no possibility to arrive at a chiral anomaly. However, for certain relations between the LV parameters, the divergent part of the CFJ term is ruled out, and for a certain regularization, its finite part also vanishes. It should be noted that the analogous situation, that is, the complete elimination of the CFJ term for a certain relation between LV parameters, takes place for the generation of the CFJ term with the use of the dimension-5 LV operators \cite{dim51}. Also, for the first time, we succeeded in generating the five-derivative CFJ-like term, generalizing the results obtained earlier in \cite{Leite:2013pca}, where the three-derivative CFJ-like term has been generated. We note, however, that within our scheme, unlike that one used in \cite{Leite:2013pca}, the three-derivative CFJ-like term does not arise.

Some possible next steps in the development of our study could consist, first, of a detailed study of possible higher-derivative (especially, five-derivative) generalizations of the Myers-Pospelov term \cite{Myers:2003fd}, including their impact within modified dispersion relations, second, of studies of LV QED with CPT-even dimension-6 terms. We expect to perform these studies in forthcoming papers.

{\bf Acknowledgments.}  The work of T. M. has been partially supported by the CNPq project No. 309360/2025-0 and FAPEAL project No. E:60030.0000002341/2022. The work of A. Yu.\ P. has been partially supported by the CNPq project No. 303777/2023-0.

\end{document}